\documentclass[agupp]{aguplus}
\usepackage{graphics}

\lefthead{Baumert}
\righthead{Primitive turbulence}

\authoraddr{H. Z. Baumert, IAMARIS, Bei den M\"uhren 69A,
D-20457 Hamburg, Germany (baumert@iamaris.de)}

\usepackage{graphicx}
\graphicspath{{converted_graphics/}}
\begin{document}
\bibliographystyle{ametsoc}

\def\lesssim{\mathrel{\hbox{\rlap{\hbox{\lower0.45em\hbox{$\sim$}}}\hbox{$<$}}}}
\def\gtrsim{\mathrel{\hbox{\rlap{\hbox{\lower0.45em\hbox{$\sim$}}}\hbox{$>$}}}}
\def\gae{\mathrel{\hbox{\rlap{\hbox{\lower0.45em\hbox{$\sim$}}}\hbox{$>$}}}}
\def\lae{\mathrel{\hbox{\rlap{\hbox{\lower0.45em\hbox{$\sim$}}}\hbox{$<$}}}}

\title{Primitive turbulence: kinetics, Prandtl's\\ mixing
length, and von K{\'a}rm{\'a}n's constant
}

\author{Helmut Z.\ Baumert}
\affil{Freie Universit\"at Berlin, Germany\\ Dept.\ Mathematics and Computer Science\\ 
}

\begin{abstract}
\noindent
The paper presents a  theory of shear-generated turbulence at asymptotically high Reynolds numbers.
It is based on an ensemble of dipole vortex tubes taken as quasi-particles and realized in form of
rings, hairpins or filament couples of potentially finite length. In a not necesserily planar cross 
sectional area through a vortex tangle, taken locally orthogonal through each individual tube, 
the dipoles are moving  with the classical dipole velocity $u=r\cdot \omega$. The vortex radius $r$ is directly
related with Prandtl's classical mixing length. The quasi-particles perform dipol chaos which reminds of 
molecular chaos in real gases. Collisions between quasi-particles lead either to particle annihilation (turbulent dissipation) 
or to particle scattering (turbulent diffusion). These ideas suffice to develop a closed   
theory of shear-generated turbulence without empirical parameters, with analogies to birth 
and death processes of macromolecules. It coincides almost perfectly with the well-known 
$K$-$\Omega$ turbulence closure applied in many branches of science and technology. 
In the case of free homogeneous decay the TKE is shown to follow $t^{-1}$. For an adiabatic 
condition at a solid wall the theory predicts a logarithmic mean-flow boundary layer with von 
K{\'a}rm{\'a}n's constant as $(2\,\pi)^{-1/2}\approx 0.399$.

\begin{flushright}
\textit{``Everything should be made as simple as possible, but not simpler.''} \\
 -- Albert Einstein
\end{flushright}

\vspace{0.25cm}

\end{abstract}


\section{Introduction\label{intro}}  
          
\noindent          
$\texttt{"}$The diversity of problems in turbulence should not obscure the fact 
that the heart of the subject belongs to physics.$\texttt{"}$ (Falkowski and Sreenivasan 2006)\nocite{falkovichsreenivasan06}.
Most classical turbulence theories and models rest more or less on the Fridman-Keller (1924)
\nocite{kellerfridman24} series expansion of the Navier-Stokes equation, the first element of which is the 
Reynolds (1895) equation. Higher elements are the subject of turbulence closure models (e.g.\ Wilcox 2006). 
\nocite{reynolds1895} This approach attracted many researchers, but 
could not solve most elementary problems of turbulence. On the contrary, it led to more and 
more `universal' higher-order equations. Although also the writer of this article made his first 
steps towards turbulence along this path, he became more and more sceptic that it is the right one 
to proceed. Therefore, with the exception of the mean-flow kinetic energy balance, the 
following text does not take much advantage from the Fridman-Keller expansion.

Below a new path is testet. It is mathematiclly simple, and for some readers
even primitive so that the writer felt obliged to announce this already 
in the headline. Nevertheless, a primitive model sometimes highlights 
aspects which are overseen in more elaborate theories (Lorenz 1960). 

The following development explores physical analogies between turbulence 
taken as an ensemble of vortex-dipol tubes and real gases of semi-stable macromolecules. 
The underlaying non-linear statistical theory of macromolecular gas kinetics is not 
repeated here and can be found, for example, in textbooks on synergetics (Haken 1978) \nocite{haken78,haken83}.

While in ideal gases molecules are point masses with zero cross section and 
infinite free path, in real gases the cross sections are finite so that molecule-molecule 
collisions limit the free path. A collision of two molecules (higher-order collisions 
are neglected here for simplicity) can have one out of two possible results: 
\begin{itemize}
  \item [(i)] The molecules 
	are spatially scattered, which corresponds to molecular diffusion and mixing. 
  \item[(ii)] The molecules vanish due to their semi-stability, which corresponds
	to particle annihilation or concentration decay. 
\end{itemize}
The turbulent analogues of the above two processes are (i) turbulent diffusion
and (ii) annihilation of vortices or dissipation of turbulent kinetic energy. The particle dynamics
(molecules, vortices) can be described by the following equation, which is a special case of the 
Oregonator (a certain class of diffusion-reaction systems, see e.g.\ Ch.\ 9 in Haken, 1978): 
\begin{eqnarray}
\label{n-1}
	\frac{\partial n}{\partial t}&=& 
	 \frac{\partial}{\partial \vec x}
	\left(\nu\, \frac{\partial n}{\partial \vec x} \right) - \beta\,n^2 \,.
\end{eqnarray}
Here $n$ is the particles' volume density, $\nu$ is the diffusivity, and $\beta$ is a constant. 

Equation (\ref{n-1}) describes the joint effect of two irreversible processes: 
particle diffusion through 
${\partial}/{\partial \vec x}\left(\nu\, {\partial n}/{\partial \vec x} \right)$,
and irreversible decay through $-\beta\,n^2$. It forms an initial-value problem 
for $n=n(\vec x, t)$. The final state is $n(\vec x, \infty)=0$. Finite molecule numbers 
can only be kept when new particles are continuously added to the volume under study. 

The following theory uses discrete vortex dipoles
instead of the a.m.\ semi-stable macromolecules. They are quasi-particles, i.e.\  
locally excited states of an otherwise homogeneous fluid like phonons in solid-state physics.
At very high Reynolds number $Re$ we expect that the fluid contains asymptotically 
many vortices and exhibits universal behavior in the sense of the Kolmogorov (1941) \nocite{K41}
 scaling laws. 

\section{Vortices\label{vortices}}  

\noindent
The classical closed vortex line (e.g.\ the closed centerline of an idealized smoke ring) 
represents an exact weak solution of the Euler equation, has infinitely thin diameter, 
infinitely high angular velocity, but finite circulation, $c=\pi\,r^2\,\omega< \infty$, 
where $r$ and $\omega$ are effective radius and vorticity.
The fluid outside the vortex line is inviscid and irrotational. The classical
vortex is an abstract frictionless object. 

The Batchelor couple (Lesieur 1997) \nocite{lesieur97} is a vortex dipol, consisting of 
two anti-parallel vortex lines. It is also an idealized frictionless object. 
When isolated and far from boundaries, in 3D the trajectory of a 
Batchelor couple is a straight line. The flow field of 
one vortex moves the other vortex and \textit{vice versa}. 
In practice such couples are stable over short to moderate 
propagation times. They conserve their kinetic 
energy, circulation (which is zero for this couple by definition), 
their vortex radii and vorticities (which carry opposite signs). 
The Batchelor couple propagates into the same direction as
the fluid moves between the two vortices forming the couple. 
In a sufficiently dense ensemble of Batchelor couples their trajectories are no longer
straight lines due to mutual interactions.

The counterpart of the Batchelor couple will be called below a von-K\'{a}rm\'{a}n couple.
It is a pair of parallel vortex lines of non-zero total circulation.
In an initial phase of their evolution they rotate around their common center of gravity,
which remains in rest. Such a couple is known since long to be fundamentally unstable.
Its kinetic energy is dissipated into heat (e.g.\ Lamb 1932) \nocite{lamb32}. 

\subsection{Vortex dipoles $vs.$ vorticons\label{sub_vorticons}}

\noindent
A vorticon is a narrow relative
of the Batchelor couple, with the following differences: the effective radius $r$ 
and its kinetic energy $u^2/2=r^2\omega^2/2$ are finite. These 
conditions are well realized in practice, at best in quantum turbulence 
(cf. Vinen and Donelly 2007) \nocite{vinendonelly07}. \\

\nocite{maxwell1867,einstein05a}

\noindent
{\bf Vorticon generation = turbulence production:} Due to the  
conservation principle for circulation in ideal fluids (Helmholtz 1858), 
in a circulation-free volume vortices can be generated (and annihilated) 
only in form of anti-parallel vortex pairs with vanishing circulation. 
I.e. turbulence kinetic energy (TKE) generation by shear 
is generation of quasi-particles in the above sense.\\

\noindent
{\bf Vorticon annihilation = TKE dissipation:}  Vorticons can be annihilated in collisions
if the two collision partners are reorganized into von-K\'{a}rm\'{a}n couples.
Consider two initial vorticons, $(+\Uparrow-)^1$ and $(-\Downarrow+)^2$, 
which reorganize during collision into the following von-K\'{a}rm\'{a}n  couples: 
($+^1\;\|\;+^2$) and ($-^1\;\|\;-^2$), where the numbers $^1$ and $^2$ refer to 
the numbered particles. In both new `particles' the non-zero molecular viscosity 
of a real fluid leads to kinetic energy dissipation and finally to a merger of the 
like-signed vortices on a lower energy level (see also Klein \& Majda 1993). This `inelastic' 
collision happens when the angle $\varphi$ between the propagation 
directions of the two colliding vorticons belongs to the interval 
$\frac{1}{2}\, \pi \le \varphi < \frac{3}{2}\,\pi$ (forward facing half sphere). 

In the initial definition of our quasi-particles we have neglected
viscosity. We re-introduce this important property 
by assuming that von-K\'{a}rm\'{a}n couples are absolutely instable 
and instantaneously converted into heat. 

\noindent
{\bf Vorticon scattering = turbulent diffusion:}  Vorticons are scattered
in collisions between two vorticons if the two are reorganized into 
two new Batchelor couples with new propagation angles. 
Consider again two initial vorticons, 
$(+\Uparrow-)^1$ and $(-\Downarrow+)^2$. After a collision 
they may form a new quasi-particle with a new propagation velocity vector:
($+^1\;\|\;-^2$) and ($-^1\;\|\;+^2$). This  `elastic' collistion happens 
when the angle $\varphi$ between the propagation directions of the 
two colliding vorticons belongs to the backward facing half sphere. 

\subsection{Vorticon tangle\label{tangle}}

\noindent
At high $Re$ vorticons are assumed to form a dense three-dimensional tangle.
As already found for 2D trajectories (Aref 1983, Eckhardt and Aref 1988) \nocite{aref83,eckhardtaref88}, 
in 3D vorticons as elements of a tangle propagate along complex non-linear trajectories 
until collision with other vorticons. But we expect that the tangle is so dense 
that the effective trajectories are short and can be treated as straight lines.
Frisch (1995) \nocite{frisch95} suggested that  
``The idea of conservative dynamics punctuated by dissipative 
events could be directly relevant in three dimensions.''
This is exactly what is elaborated here. 
Our form of `conservative dynamics' is chaotic dipol motion which conserves
energy until particle annihilation. The trajectories are `punctuated' by 
dissipative collisions which lead to vorticon conversion into heat.

\begin{figure}[htb]
\centerline{\includegraphics[width=6cm,height=7.72cm,keepaspectratio]{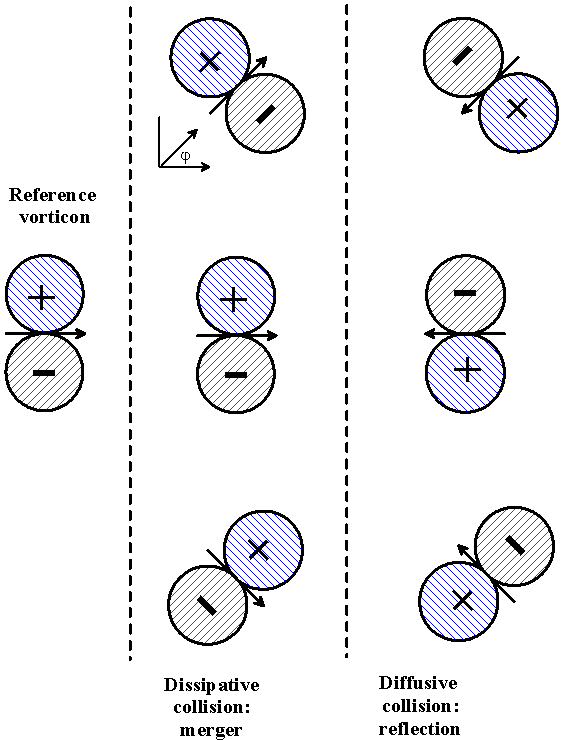}}
  \caption{Two elementary vorticon-vorticon 
	interactions. On the very left the reference vorticon is shown
	which is assumed to move to the right.
	The middle column shows different orientations of the target vorticon
	which all lead to a dissipative (inelastic) collision, i.e.\ to annihilation of a 
	quasi-prarticle. The right column shows potential target vorticons 
	in elastic scatter position corresponding to turbulent diffusion.
	A collision with these targets conserves energy. 
	\label{f:vorticons-free}}
\end{figure}

Figure \ref{f:vorticons-free} exhibits a cross section through 
vorticons. The arrows denote the propagation direction. 
A plus sign indicates rotation against the clock.

\subsection{Historical notes\label{historical}}

\noindent
Whereas the present approach to turbulence is based on ensembles of $discrete$ quasi-particles,
most turbulence closures of the past take the fluid as a $continuum$.
Both concepts are old. James Clerk Maxwell (1867) gave a detailed
discussion of this dichotomy in connection with his kinetic theory 
of gases and traced it back to Democrit and Lucretius. 
Ren{\'e} Descartes (1596 -- 1650) spoke about `tourbillons' forming the universe;
based on the results of Herrmann von Helmholtz (1858)
Lord Kelvin (1867a, b) coined the notion `vortex atoms'; 
for a comprehensive overview see (Saffman 1992) \nocite{saffman92}.
\nocite{descartes1644}

Most of later analytic efforts towards a kinetic theory of vortices were restricted 
to spatially two-dimensional cases wherein the Hamilton formalism is applicable 
(Onsager 1949) \nocite{onsager49}. Exceptions are vortex-based numerical simulation receipes and rules.

Marmanis (1998) \nocite{marmanis98} was the first who proposed 
the vortex dipole as $the$ fundamental quasi-particle of turbulence. He wrote:
$\texttt{"}$The introduction of the \textit{vortex dipole chaos assumption} permits 
one to derive a kinetic equation for a `gas' of vortex dipoles.$\texttt{"}$ 
He presented a methodical mixture of Onsager's theory with Bogoljubov's 
perturbation-kinetic approach with application to 2D inviscid turbulence.
Dissipative 3D turbulence has not been treated by this author. 

Lions and Majda (2000) \nocite{lionsmajda00} tried to develop a kinetic theory of 3D turbulence at high $Re$.
They used systematic asymptotic expansions of the Navier-Stokes equation
in the Fridman-Keller sense. On the one hand they tried to overcome limitations of the 
Klein-Majda theory (Klein and Majda 1993) \nocite{kleinmajda93}, on the other hand they limited 
themselves to the \textit{quasi-parallel} case where the Klein-Majda theory is 
constructively applicable through a rigorous equilibrium statistical formalism.

Chen et al. 2004 \nocite{chen-etal04} also aimed at the development of a physico-kinetic turbulence theory. 
They explicitely envisaged Boltzmann's approach and developed a complex
formalism, but still without practically  applicable predictions. 

Despite high efforts, these works did not yet lead to unique and constructive rules for the 
practical computation of the eddy viscosity from mean flow data. Unfortunately, 
with a few exceptions eddy viscosity is the most important turbulence parameter 
in computational aero-, hydro- and thermo-dynamics. 

Below turbulence balance equations are derived which describe the local-average 
dynamics of  turbulence kinetic energy, $K$,  mean vorticity, $\omega$, and eventually eddy 
viscosity, $\nu$. These equations are only a specific format or a specific implementation of the 
primitive theory. In special cases they can be solved analytically. 

In more general cases one would try to solve them using finite-difference or 
finite-volume etc. methods. If the solution shall be approximated by the  
Monte-Carlo method then the writer recommends to start not from the 
continuum equations derived below, but better directly from the mechanical
picture of the vorticon ensemble, preferably using the already 
existing eddy-collision methods where various vortex-filament primitives 
are already avaialable (Cottet and Koumoutsakos 2000, Andeme 2008)\nocite{cottet-koumoutsakos00,andeme08}.

\section{Primary variables in primitive turbulence\label{primvars}}

\subsection{TKE and vorticity\label{tkeandvorticity}}

\noindent
For the construction of the eddy viscosity, which is proportional to TKE $times$ unit time,
and of the TKE dissipation rate, which is TKE  $per$ unit time, two constituing variables are necessary: TKE and 
a measure of time. Here the below-defined quantities $K$ and $\omega$ are chosen. They are introduced now.

Consider a small volume element $\delta V$ populated by an ensemble of $j = 1 \dots n$ 
dipoles with individual effective vortex radii $r_j$ and vorticities $\omega_j$. During the free  
flight time $\tau_j$ the properties of the quasi-particle $j$ are conserved. 
The particle's volume density is $n/\delta\cal V$.
The total TKE within $\delta \cal V$ is the sum of the kinetic energies
of the individual vorticons:
\begin{eqnarray}
	K_{\delta \cal V}&=& 
	\frac{1}{\delta\cal V}
	\sum_{j\epsilon\delta\cal V} 
	\frac{1}{2}\,r_j^2\,\omega_j^2\;
	=\;\frac{n}{\delta \cal V}\;\bar k\;\label{Knk-1},\\
	\omega_{\delta \cal V}&=& \frac{1}{\delta\cal V}\sum_{j\epsilon\delta\cal V}|\omega_j|
	\;=\;\frac{n}{\delta \cal V}\;\bar\omega\,.\label{omeganom-1}
\end{eqnarray}
Multiplication of (\ref{Knk-1}, \ref{omeganom-1}) with $\delta\cal V$ gives 
\begin{eqnarray}
	K\;=\; \delta \cal V\cdot K_{\delta \cal V}&=& 
	\sum_{j\epsilon\delta\cal V} 
	\frac{1}{2}\,r_j^2\,\omega_j^2\;
	=\;n\;\bar k\;\label{Knk-2},\\
	\omega\;=\; \delta \cal V\cdot 
	\omega_{\delta \cal V}&=& \sum_{j\epsilon\delta\cal V}|\omega_j|
	\;=\;{n}\;\bar\omega\,.\label{omeganom-2}
\end{eqnarray}
$K_{\delta \cal V}$, $\omega_{\delta \cal V}$ are local volume densities 
of TKE and vorticity magnitude, respectively. They and  $K,$ $\omega$ 
are extensive variables by definition, i.e.\ they 
change when new particles with average properties $\bar k$, $\bar\omega$ 
are added to $\delta\cal V$, as then $K \rightarrow(n+1)\,\bar k$ 
and $\omega\rightarrow(n+1)\,\bar\omega$. 

$\bar k$ and $\bar \omega$ are ensemble averages and as such 
intensive variables which do not change when new particles with 
average properties are added to the ensemble in $\delta \cal V$:
\begin{eqnarray}
	\bar k&=& \frac{1}{n}
	\sum_{j\epsilon\delta\cal V} \label{Knk-3}
	\frac{1}{2}\,r_j^2\,\omega_j^2\,,\\
	\bar \omega&=& \frac{1}{n} \sum_{j\epsilon\delta\cal V}|\omega_j|\,.\label{omeganom-3}
\end{eqnarray}

\subsection{Auxiliary quantities\label{aux}}

\noindent
We introduce useful auxiliary quantities: 
the ordinary vorticity frequency, $\Omega=1/T$ 
and the often used constant $\kappa$:
\begin{eqnarray}
	\omega &=& 2\,\pi\,\Omega \;=\;\Omega/\kappa^2\,\label{Omega},\\
	\kappa &=& (2\,\pi)^{-1/2}\approx \,0.399 \,\label{kappa1}\,.
\end{eqnarray}
Further below, $\kappa$ appears to be von K\'{a}rm\'{a}n's constant.
While $|\omega_j|=2\pi/T_j$ is an angular frequency, $\Omega$ is an ordinary frequency. 

\subsection{Derived variables\label{derived}}

\noindent
{\bf Effective radius $R$.} $R^2$ is defined here as the weighted ensemble mean:
\begin{eqnarray}
	R^2 &=& \frac{\sum_{j\epsilon\delta\cal V}\;r_j^2\,\omega_j^2}
	{(\sum_{j\epsilon\delta\cal V}\;|\omega_j|)^2} 
	\;=\; \frac{2\,K}{\omega^2}
	\;=\;\frac{2\,\bar k}{\bar\omega^2}\; n^{-1}\label{R2-1}\,.
\end{eqnarray}
It depends (inversely) on $n$ and is thus an extensive variable.
In freely decaying homogeneous turbulence (Section \ref{homogeneous}) only the particle number decreases
while the ensemble mean properties $\bar k$ and $\bar \omega$ remain constant in time such that 
\begin{eqnarray}
	n\, \pi \, R^2 &=& \mbox{const.} \label{compressibility} 
\end{eqnarray}
This reminds of a 2D version of the equation of state of a real gas telling us 
that the (planarized) locally orthogonal cross sectional area through a dense 
vorticon tangle is compactly filled with dipoles.

\noindent
{\bf Dissipation rate.} The rate of TKE dissipation,
$\varepsilon$, has the units of TKE per time [m$^2$ s$^{-3}$]. Up to a factor 
$\zeta$ it is governed by the product $K\cdot \Omega$, :
\begin{eqnarray}
	\varepsilon &=& \zeta \,\frac{K\,\Omega}{\pi}
	\;=\; \zeta \,\frac{\bar k\,\bar \omega}{\pi}\, n^{2}
	\,\label{eps1}.
\end{eqnarray}
$\varepsilon$ is an extensive variable. 
Below we show that $\zeta\equiv 1$. In freely decaying homogeneous
turbulence (\ref{eps1})  is identical with the quadratic term in
(\ref{n-1}) such that in this case 
\begin{equation}\label{inthiscase}
	\beta = \zeta\,\bar k\,\bar\omega/\pi\;=\; \mbox{const.}
\end{equation}

\nocite{prandtl25}
\noindent 
{\bf Eddy viscosity.} Ludwig Prandtl (1925) noticed two simultaneous properties
of his mixing length, namely that it (a) ``\dots may be considered as 
the diameter of the masses of fluid moving as a whole in each individual case; or again, 
as the distance traversed by a mass of this type before it becomes blended 
in with neighboring masses \dots''; and also that it is (b) ``\dots somewhat similar, 
as regards effect, to the mean free path in the kinetic theory of gases \dots'' 
(Bradshaw 1974)\nocite{bradshaw74}. 

We interpret Prandtl's mxing length in terms of `our' $R$ in (\ref{R2-1}) because 
it is proportional to (a) the local mean radius of vorticons and, simultaneously, (b) to the free path.  

\begin{figure}[tbp] 
\centerline{\includegraphics[width=6cm,height=5.81cm,keepaspectratio]{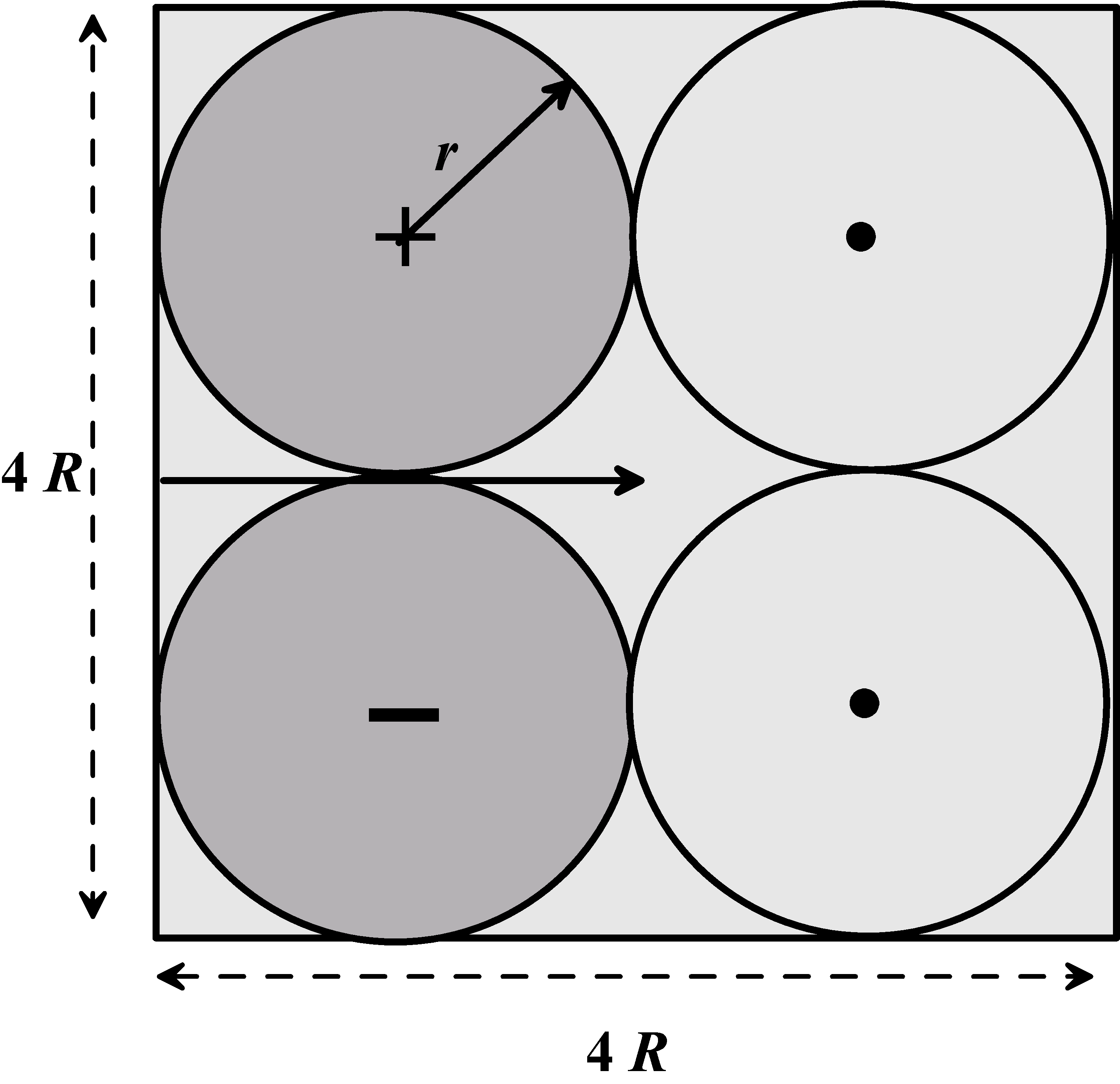}}
  \caption{Local cross section through a dipol  tangle  far from boundaries
with maximum possible density at $Re\rightarrow\infty$. 
The dark grey vorticon in the left half square labelles the instantaneous
situation, the `empty vorticon' in the right half square indicates the vorticon
after its free flight from the left to the right into its collision position, where
it is either annihilated or scattered. 
Far from boundaries a vorticon obviously `needs' a cross sectional area of 
${\cal A}_{free} = (4\,R)^2$).}
\label{f:vorticons-area-1}
\end{figure}

To explore this idea further we follow Albert Einstein's (1905) theory of molecular diffusion
in fluids and use his expression for the coefficient of diffusion now for our 
coefficient of turbulent diffusion, or eddy viscosity $\nu$, in terms of a mean free path, 
$\lambda$, and a mean free flight time, $\tau$:
\begin{eqnarray}
	\nu &=& {\lambda^2}/{2 \,\tau}\,\label{nu-1}\,.
\end{eqnarray}
For a vortex dipole these parameters can directly be given as follows 
wherein $u$ is the mean (linear) advection velocity of a vortex dipole:
\begin{eqnarray}
	\tau&=& {\lambda}/{u}\,\label{tau},\\
	u &=& \sqrt{2\,K} \,.\label{u}
\end{eqnarray}
For a quasi-steady turbulent vorticon tangle at $Re=\infty$ and far from solid boundaries 
the mean free path will clearly be very limited but it cannot vanish. The packing density 
should be dense but small enough that during a flight a particle can be replaced by a newly 
generated. This implies that \nocite{baumert05b} (Baumert 2005b)
\begin{eqnarray}
	\lambda &=& 2\,R \,\label{lambda-2}.
\end{eqnarray}
The free area of a free vorticon far from boundaries is (Fig.\ \ref{f:vorticons-area-1})
\begin{eqnarray}
	{\cal A}_{free} &=& (4\,R)^2\,.\label{vorticons-area-free}
\end{eqnarray}
We finally get the eddy viscosity as an intensive variable as follows:
\begin{eqnarray}
	\nu &=& K/\pi \, \Omega \;=\; 2\,K/\omega \;=\;\omega\; R^2\,,\label{nu-2}
\end{eqnarray}
This equation is called the Prandtl-Kolmogorov relation. 

\section{Free homogeneous decay of turbulence\label{homogeneous}}

\noindent
For constant properties $\bar k$, $\bar\omega$ of our quasi-particles, following
(\ref{Knk-3}, \ref{omeganom-3}) the variables $K$ and $\Omega$ are proportional to 
the vorticon density $n$ and have therefore to satisfy the same diffusion-reaction 
equation (\ref{n-1}) like $n$, because these properties are `solidly mounted' at the 
dipoles. In the case of decay we have thus the following:
\begin{equation}
	n\;=\; \frac{K}{\bar k} \;=\;\frac{\omega}{\bar\omega}\,.
\end{equation}
In analogy to (\ref{n-1}) it holds that 
\begin{eqnarray}
\label{tke-0}
	\frac{dK}{dt}+\beta_K\,K^2&=& 0\;,\\
\label{om-0}
	\frac{d\Omega}{dt}+\beta_{\Omega}\,\Omega^2&=& 0\;.
\end{eqnarray}
The  reader easily sees that at large $t$
\begin{eqnarray}
\label{tke-1}
	K(t)& = & (\beta_K\,t)^{-1}\;,\\
\label{om-1}
	\Omega(t)&=& (\beta_{\Omega}\,t)^{-1}\;,
\end{eqnarray}
and 
\begin{eqnarray}
\label{nu-3}
	\nu(t) &=& \frac{K(t)}{\pi\,\Omega(t)}
	\;=\;\frac{\beta_{\Omega}}{\pi\,\beta_K}\;=\;\mbox{const.}
\end{eqnarray}
Equation (\ref{tke-1}) coincides with the results of a fairly general 
similarity analyses of the Navier-Stokes equations by Oberlack (2002) \nocite{oberlack02}
and with the experimental results of Dickey and Mellor (1980)\nocite{dickeymellor80}.
\nocite{baumert05a,baumert05b}


\section{Forced inhomogeneous turbulence\label{forced}}

\noindent
In contrast to Section \ref{homogeneous} now the assumptions 
of free decay and spatial homogeneity are given up. This implies that we need to add source 
($F_K$ and $F_{\Omega}$) and diffusion terms to the right-hand sides of equations (\ref{tke-0}, \ref{om-0}):
\begin{eqnarray}
\label{K-1}
	\frac{\partial K}{\partial t}+  \beta_K\,K^2 &=&  F_K
	+\frac{\partial}{\partial \vec x}
	\left(\nu\, \frac{\partial K}{\partial \vec x} \right) ,\\
\label{Om-1}
	\frac{\partial \Omega}{\partial t}+\;\beta_{\Omega}\,\Omega^2 &=&F_{\Omega}
	+ \frac{\partial}{\partial \vec x}
	\left(\nu\,\frac{\partial\Omega}{\partial\vec x}\right)\,.
\end{eqnarray}
The factors $\beta_K$ and $\beta_{\Omega}$ are not longer expected to be constants. 

\subsection{TKE production by mean-flow shear: $F_K$\label{tkeprod}}

\noindent
The specification of the TKE production term $F_K$ for a shear flow 
can be obtained from textbooks (Wilcox 2006, Schlichting and Gersten 2000)\nocite{wilcox06,schlichting-gersten00},
\begin{equation}\label{tke-production}
	F_K \;=\; \nu \, S^2 \;=\; \frac{K}{\pi\,\Omega} \,S^2\,,
\end{equation}
where $S^2$ is the total instantaneous shear, squared,
\begin{equation}\label{def-shear}
	S^2 \; = \; \sum_{i,j=1}^3 \left( 
	\frac{\partial U_j}{\partial x_i}+
	\frac{\partial U_i}{\partial x_j} \right)\;
	\frac{\partial U_i}{\partial x_j}\,,
\end{equation}
and $U_i$ is the $i^{th}$ component of the mean flow velocity vector $(U_1, U_2, U_3)^T$. 

\subsection{Vorticity generation by mean-flow shear: $F_{\Omega}$\label{vortgen}}

\noindent
The above formula (\ref{tke-production}) for $dK/dt=F_K$ is classical hydrodynamic knowledge. However, 
the specification of the vorticity generation term $F_{\Omega}$ is less trivial. Here one
has to consider a fundamental argument of Henry Tennekes. It has been carefully 
discussed in the past, see Tennekes 1989; Baumert and Peters 2000, 2004; 
Kantha 2004, 2005; Kantha et al.\ 2005; Kantha and Clayson 2007. We cast Tennekes' argument into mathematical form.
He hypothesized that, on dimensional grounds, the length scale (here: $L$ or $R$) cannot depend on the ambient
shear for a neutrally stratified homogeneous shear flow. Since shear production $F_K$ involves shear, $F_{\Omega}$
needs to be constructed such that the role of shear vanishes in the evolution equation for the length scale. The latter is
generally derived from equation (\ref{R2-1}) as 
\begin{equation}\label{tennekes-1}
	\frac{1}{R^2}\,\frac{dR^2}{dt}\;=\;\frac{1}{K} \,\frac{dK}{dt}-2\,\frac{1}{\omega}\,\frac{d\omega}{dt}\,.
\end{equation}
We insert $dK/dt$ from (\ref{tke-production}) in (\ref{tennekes-1}) and find
\begin{equation}\label{tennekes-2}
	\frac{1}{R^2}\,\frac{dR^2}{dt}\;=\;\frac{S^2}{\pi\,\Omega} - 2\,\frac{1}{\omega}\,\frac{d\omega}{dt}\,.
\end{equation}
In terms of the present theory, Tennekes' hypothesis means $dR^2/dt=0$, which implies that 
\begin{equation}\label{tennekes-3}
	\frac{S^2}{\pi\,\Omega} \;=\; 2\,\frac{1}{\omega}\,\frac{d\omega}{dt}\,.
\end{equation}
It is left to the reader to see that (\ref{tennekes-3}) gives
\begin{eqnarray}
	\label{def-omega-1}
		 \frac{d\Omega}{dt}&=& \frac{S^2}{2\,\pi}\,.
\end{eqnarray}
As far as  (\ref{R2-1}) is a pure ensemble-mean definition which considers 
neither diffusion nor annihilation of particles, the time derivative in (\ref{def-omega-1}) 
needs to be understood as the pure generation term in (\ref{Om-1}), $F_{\Omega}$:
\begin{eqnarray}
	\label{def-FOmega-1}
		 F_{\Omega}&=& \frac{S^2}{2\,\pi}\,.
\end{eqnarray}

\subsection{The multipliers $\beta_K$ and $\beta_{\Omega}$ \label{multipliers}}

\noindent
Let us summarize (\ref{K-1}, \ref{Om-1}) and (\ref{tke-production}, \ref {def-FOmega-1}) as follows,
\begin{eqnarray}
\label{K-3}
	\frac{\partial K}{\partial t}
	-\frac{\partial}{\partial \vec x}
	\left(\nu\, \frac{\partial K}{\partial \vec x} \right) &=&  \frac{K}{\pi \,\Omega}\,S^2 -  \beta_K\,K^2 \, ,\\
\label{Om-3}
	\frac{\partial \Omega}{\partial t}
	-\frac{\partial}{\partial \vec x}
	\left(\nu\,\frac{\partial\Omega}{\partial\vec x}\right)
	&=&\frac{S^2}{2 \pi} -\beta_{\Omega}\,\Omega^2\,.
\end{eqnarray}
where $\nu$ is known as function of $K$ and $\Omega$ through (\ref{nu-2}).
It remains to determine the still unknown multipliers $\beta_K$ and $\beta_{\Omega}$.
This can be accomplished as follows. 

We note that the term $\beta_K\,K^2$ in (\ref{K-3}) 
is identical with the dissipation rate (\ref{eps1}) of TKE:
\begin{eqnarray}
\label{K-2}
	\varepsilon &=&  \zeta \,\frac{K\,\Omega}{\pi}\;=\; \beta_K\,K^2  \,.
\end{eqnarray}
From the second equation in (\ref{K-2}) we conclude with (\ref{nu-2}) that
\begin{equation}\label{K-22}
	\beta_K \;=\; \frac{\zeta}{\nu\,\pi^2}\,,
\end{equation}
while from (\ref{nu-3}) it follows that
\begin{equation}\label{K-23}
	\beta_{\Omega} \;=\; \nu\,\pi\,\beta_K\,.
\end{equation}
We insert (\ref{K-22}) into (\ref{K-23}) and see that
\begin{equation}\label{K-24}
	\beta_{\Omega} \;=\; \frac{\zeta}{\pi}\,.
\end{equation}
Inserting (\ref{K-22}, \ref{K-24}) in (\ref{K-3}, \ref{Om-3}) we get, after some algebra using (\ref{nu-2}), 
an almost complete system wherein only $\zeta$ is still to be specified:
\begin{eqnarray}
\label{K-4}
	\frac{\partial K}{\partial t}
	-\frac{\partial}{\partial \vec x}
	\left(\nu\, \frac{\partial K}{\partial \vec x} \right) &=&  \nu\left(S^2 -  \zeta \,\Omega^2\right) ,\\
\label{Om-4}
	\frac{\partial \Omega}{\partial t}\;
	-\frac{\partial}{\partial \vec x}
	\left(\nu\,\frac{\partial\Omega}{\partial\vec x}\right)
	&=&\frac{1}{\pi}\left( \frac{1}{2}\,S^2 -\zeta \,\Omega^2\right).
\end{eqnarray}

\section{Turbulent boundary layer\label{boundlayer}}

\noindent
Consider a stationary boundary layer close to a plane solid wall at $z = 0$ where
$z$ is the only coordinate of interest here. It points orthogonal from the wall into the fluid.
The mean flow is parallel to the wall, i.e.\ $U_1=U>0$ and $U_2=U_3=0$. Consequently
with (\ref{def-shear}) we have to write
\begin{equation}\label{newshear}
	S \;=\; \left|dU/dz\right|.
\end{equation}
The diffusive TKE flux into the viscous sublayer at $z=z_0\rightarrow 0$ has  to vanish,
\begin{equation}\label{K-25}
	\left( \nu \,dK/dz\right)_{z=z_0} \;=\; 0\,,
\end{equation}
or
\begin{equation}\label{K-251}
	K(z) \;=\; K_0\;=\; \mbox{const.}
\end{equation}
such that in the stationary case (\ref{K-4}) gives
\begin{equation}\label{K-26}
	\Omega  \;=\; S/\sqrt{\zeta}\,.
\end{equation}

\subsection{Logarithmic law of the wall\label{loglaw}}

\noindent
We insert (\ref{K-26}) into the stationary form of (\ref{Om-4}) so that we 
have to solve the following equation for $S=S(z)$: 
\begin{equation}\label{Om-5}
	\frac{2\,K_0}{2\,\zeta-1}\,\frac{d}{dz}\left( \frac{1}{S}\,\frac{dS}{dz}\right)  \;=\; S^2\,.
\end{equation}
The solution is 
\begin{equation}\label{Om-6}
	S(z)\;=\; \frac{dU}{dz}\;=\; \frac{\sqrt{2\,K_0/(2\,\zeta-1)}}{z}\,.
\end{equation}
Integration of (\ref{Om-6}) gives the logarithmic law of the wall. 
In boundary layer theory the bottom shear stress is 
mostly defined in terms of the squared friction velocity, $u_f^2$,
\begin{equation}\label{bss}
	u_f^2 \;=\; \nu\, \frac{dU}{dz}\;=\; \frac{K_0}{\pi \,\Omega}\,S\,,
\end{equation}
and with (\ref{K-26}) it follows that
\begin{equation}\label{k-47}
	K_0 \;=\; \pi \,u_f^2 / \sqrt{\zeta}\, .
\end{equation}
This allows to rewrite (\ref{Om-6}) as follows,
\begin{equation}\label{Om-7}
	S(z)\;=\; \frac{dU}{dz}\;=\; \frac{u_f}{\tilde \kappa\,z}\,,
\end{equation}
with $\tilde\kappa$ defined through
\begin{equation}\label{Om-8}
	\tilde \kappa \;=\; \kappa \, \sqrt{(2\zeta-1)\sqrt{\zeta}} \,.
\end{equation}
Integration of (\ref{Om-7}) provides us with 
\begin{equation}\label{Om-9}
	U(z)\;=\; \frac{u_f}{\tilde \kappa}\, \ln\left(\frac{z}{z_0}\right).
\end{equation}

\subsection{Mixing length $L$\label{mixinglength}}

\begin{figure}[hbt]
\centerline{\includegraphics[width=6cm,height=7cm,keepaspectratio]{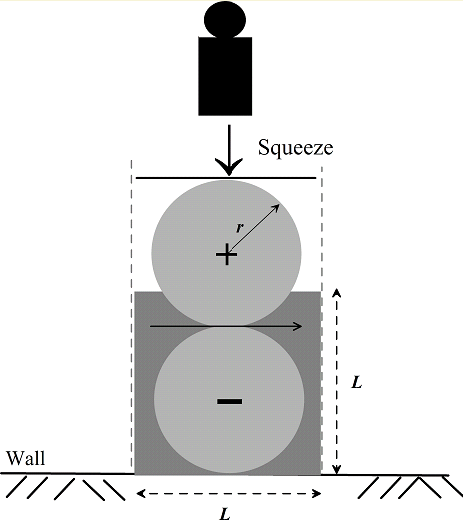}}
\caption{Cross section through a vorticon sheet at a
solid boundary. In this Figure we have $r=R$, i.e.\ the sketched 
quasi-particle is to be understood as an ensemble average. }
\label{f:vorticons-boundary}
\end{figure}

\noindent
Consider the definition of the effective vorticon radius through (\ref{R2-1}). We solve this
equation for $K$ and express the TKE in terms of $R$ and $\Omega$ as follows:
\begin{equation}\label{10}
	K\;=\; 2\,\pi^2 \,R^2\, \Omega^2\,.
\end{equation}
In practical analyses of boundary layer turbulence,  Prandtl's (1925) concept of mixing length
is still often applied. It relates the eddy viscosity to the shear. 
Following Hinze (1959, p. 279, eq. 5-2) in present notation, Prandtl defined his
mixing length $L$ as follows: 
\begin{equation}\label{11-a}
	\nu\;=\; L^2\,\left|\frac{dU}{dz}\right|\;=\;L^2\,S\,.
\end{equation}
Due to our eddy viscosity formula (\ref{nu-2}) relation (\ref{11-a}) gives 
\begin{equation}\label{11-b}
	L^2\;=\; \frac{K}{\pi\,\Omega\,S}\,,
\end{equation}
so that in the neighborhood of a solid wall we get with (\ref{K-26}) the following result:
\begin{equation}\label{11-c}
	K \;=\; \pi\, L^2\, \Omega^2\,\sqrt{\zeta}\,.
\end{equation}
Equating (\ref{10}) with (\ref{11-c}) gives 
\begin{equation}\label{11-d}
	L \;=\; \frac{R}{\kappa}\,\zeta^{\,-1/4}\, .
\end{equation}
The physical meaning of $L$ is easily understood by a glance at Fig.\ \ref{f:vorticons-boundary}.
If we may set $\zeta=1$ then (\ref{11-d}) gives together with the definition of $\kappa$ in (\ref{kappa1}) the following,
\begin{equation}
	L^2 = 2\, (\pi\,R^2)\,,\label{12-a}
\end{equation} 
which is the area of a vorticon, fully compressed into
the form of a square of side length $L$. This is the asymptotically
maximum deformation which justifies to set $\zeta \equiv 1$.


While Prandtl's mixing-length concept was applicable only in the vicinity of solid 
boundaries so that it attracted respectful criticism (e.g.\ Kundu 1990, p. 457;
Wilcox 2006), \nocite{kundu90} our concept is a generalization of Prandtl's concept and works 
also far from boundaries, even in the free stream of stratified fluids where $L$ may approach 
the Thorpe scale and/or the Ozmidov scale, depending on the conditions 
(Baumert and Peters 2004)\nocite{baumertpeters04}. 

This section shall be closed with a summary of formulae relevant at solid boundaries, 
where $z$ the distance from the wall: 
\begin{eqnarray}
	\nu&=&  u_f\,L\,\label{a-1}, \\
	L&=& \kappa\,z \,\label{a-1} ,\label{a-2}\\
	z &=& L/\kappa \;=\;R/\kappa^2 \label{a-4}\,.
\end{eqnarray}
\subsection{Boundary compression\label{compression}}

\noindent
Far from boundaries the dipole chaos is isotropic and a vorticon's
cross section exhibits circles (Fig.\ \ref{f:vorticons-free}). Closer
to a solid wall symmetry is increasingly broken 
and we see  ellipses instead. Right at the boundary
the ellipses are even deformed to an equivalent square 
(the asymptotic case) with side length 
$L=(2 \,\pi\,R^2)^{1/2}$, Fig.\ \ref{f:vorticons-boundary}.
This is the minimum area a vorticon can asymptotically attain: 
\begin{equation}\label{minimumarea}
	{\cal A}_{bound} = L^2= 2 \,\pi\,R^2\,. 
\end{equation}
Notice that the fluid within the vorticon is free of friction
so that circle, ellipse and square are energetically identical.

To estimate the degree of boundary compression of vorticons quantitatively we  
compare area (\ref{vorticons-area-free})  `occupied' by a free vorticon far from boundaries
with the corresponding area (\ref{minimumarea}) for a compressed vorticon at the very boundary:
\begin{equation}\label{compare-1}
	\frac{{\cal A}_{free}}{{\cal A}_{bound}}\;=\; \frac{(4\,R)^2}{2\,L^2}\;=\;\frac{16\,R^2}{4\,\pi\,R^2}\;=\; \frac{4}{\pi}\,\approx 1.27 \,.
\end{equation}

\begin{figure}[hbt]
\centerline{\includegraphics[width=6cm,height=4.25cm]{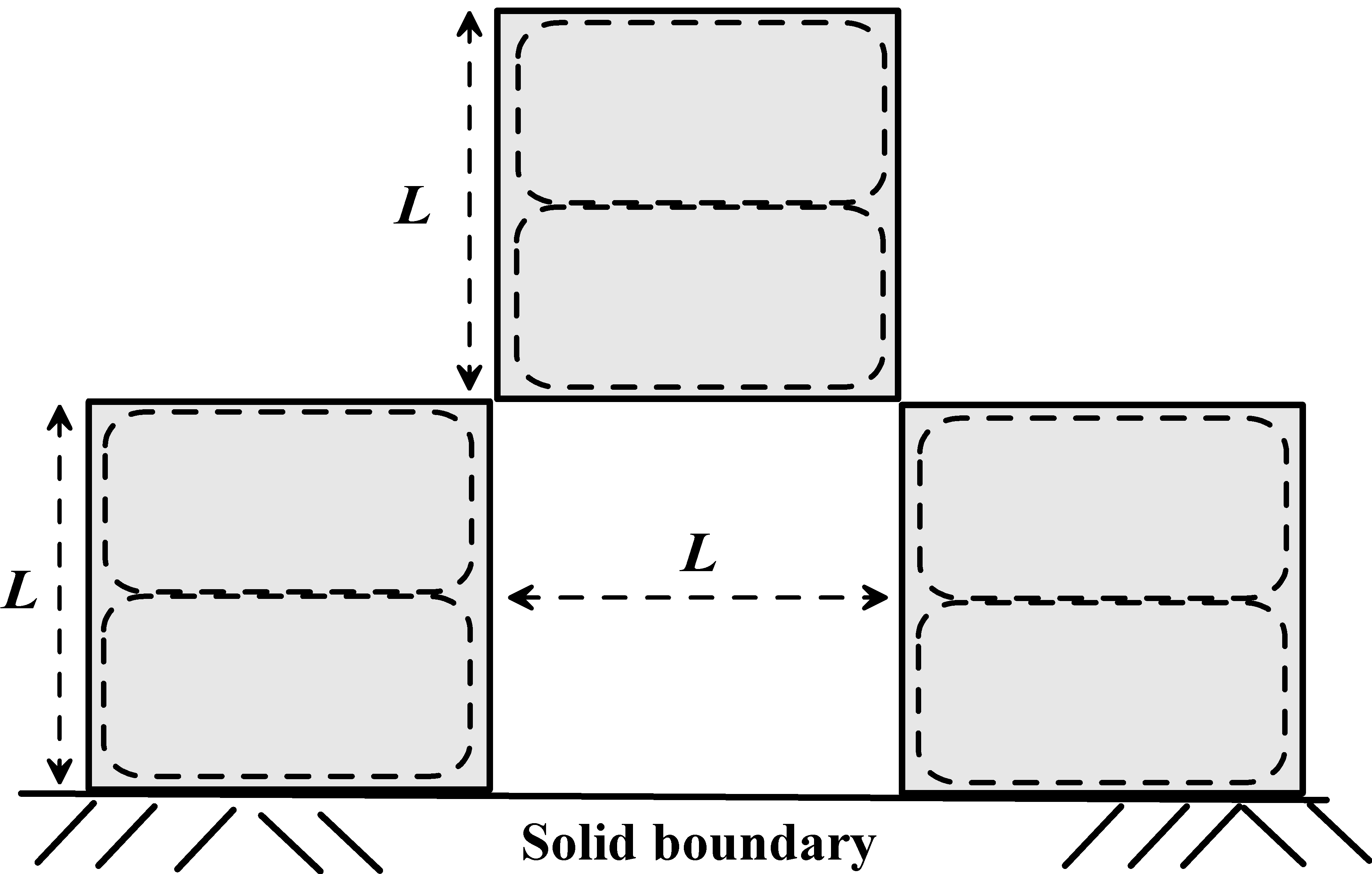}}

\caption{Squeezed vorticons like in the dark-grey square of Fig.\ \ref{f:vorticons-boundary}. 
The dotted lines symbolize the circulation pattern within the squeezed quasi-particles.}
\label{f:vorticons-area-2}
\end{figure}

\section{Discussion}

\subsection{Kinetic theory and the $k$-$\omega$ closure of Wilcox\label{KOclosure}}

\noindent
The specification $\zeta\equiv 1$ closes the primitive turbulence theory.
In compact form it reads as follows:
\begin{eqnarray}
\label{K-5}
	\frac{\partial K}{\partial t}
	-\frac{\partial}{\partial \vec x}
	\left(\nu\, \frac{\partial K}{\partial \vec x} \right) &=&  \nu\left(S^2 - \Omega^2\right),\\
\label{Om-5}
	\frac{\partial \Omega}{\partial t}\;
	-\frac{\partial}{\partial \vec x}
	\left(\nu\,\frac{\partial\Omega}{\partial\vec x}\right)
	&=&\frac{1}{\pi}\left( \frac{1}{2}\,S^2 - \Omega^2\right),\\
\label{nu-5}
	\nu&=&\frac{K}{\pi\,\Omega}\,.
\end{eqnarray}
These equations are structurally fully identical with the $k$-$\omega$ closure model discussed by 
Wilcox (2006) \nocite{wilcox06}, which is in many aspects the best out of more than 10 other semi-empirical two-equation models.
`His'  $k$ and our $K$ are equivalent.
The eddy viscosity definitions of Wilcox and the present paper agree 
effectively up to a factor of ${\cal O}(1)$.

However, `his' $\omega$ (he calls it somewhat uncertain
a `specific dissipation rate') differs significantly from our $\Omega$,
\begin{equation}
	\omega= \Omega/\sqrt{\beta^*} \approx 3.3\bar 3\cdot \Omega\,,
\end{equation}
where $\beta^*=0.09$ is one out of 6 empirical parameters of the Wilcox model.
It reproduces the logarithmic boundary layer and implies 
$\kappa=0.41$. 

A major quantitative difference between the two models 
lays in the free homogenous decay. While both exhibit $K\sim t^{-m}$, we have $m=1$, 
which agrees with Dickey and Mellor's (1980) high-$Re$ laboratory experiments
and with Oberlack's (2002) theoretical result for the free decay in the Navier-Stokes equation. 
Wilcox'  model is tuned to $m =6/5 = 1.2$, which agrees with a number of other laboratory results 
and with Oberlack's (2002) result for the free decay in the Euler equation. 

Today it is not clear
 why many decay experiments lead to $m > 1$. Possibly it is a matter of 
initial conditions, see e.g.\ Hurst and Vassilicos (2007): at high $Re$ viscosity is comparatively
small so that its regularizing effect on the initial spectrum towards a fully self-similar decay spectrum
will take more time than at lower $Re$. In some cases this time may exceed the lifetime of turbulence. 
\nocite{hurstvassilicos07}

\subsection{Von K{\'a}rm{\'a}n's number\label{vKnumber}} 

\noindent
The primitive theory makes a quantitative prediction of von K{\'a}rm{\'a}n's constant. This allows
a comparison with observations, experiments and competing other theories. 
While the latter are rare, there is a substantial literature discussing 
the precise value of $\kappa$ and its error bounds. The controversy 
whether the logarithmic or the power law is applicable to boundary layers
(Barenblatt) \nocite{barenblatt97} shall not be picked up here. The controversy on the 
universality of $\kappa$ (Chauhan et al.\ 2005)\nocite{chauhanetal05} 
is possibly only a matter of proper definition. If we restrict the 
definition of $\kappa$ to the case of $favorable$ pressure gradients only
(cf. Chauhan et al.) then we would get a unique solution which
is supported also by the Chauhan et al. data: $\kappa = 0.40$. 

A theoretical effort by Telford (1982) shall be mentioned, who derived a formula 
which connects $\kappa$ with two semi-empirical parameters, 
the entrainment and the decay constants. He predicts $\kappa=0.37$. 
\nocite{telford82,hogstrom85,telfordbusinger86,frenzenvogel94}
Further theoretical work has been done by Yakhot and Orszag (1986) \nocite{yakhotorszag86} who 
found $\kappa=0.372$ based on renormalization-group approximations.
Bergmann (1998) claimes that $\kappa=1/{\bf \it \bf e} \approx 0.368$, based on 
a new physical interpretation resting on semi-empirical considerations.

From an analysis of observational data 
H\"ogstr\"om (1985) concludes $0.40 \pm 0.01$,
which is still accepted today among most practitioners. 
But Telford and Businger (1986) draw his analysis into question and 
underline the enormous methodical problems in connection with the evaluation
of the experimental data. 

Another group of authors supports the lower values,  e.g.\ Jinyin et al.\ (2002)\nocite{jinyin-etal02}. 
They report many different $\kappa$ values obtained 
in the past from different experiments and simulations, e.g.\ back to Nikuradse, Laufer, Lawn, Perry and Abell, 
Long et al., Millikan, and Zagarola and Smits (\textit{loc.\ cit.} Jinyin \textit{et al.}). 
They summarize that the lower and upper bounds of $\kappa$  
are 0.36 and 0.45, respectively, whereby the range of experimental Reynolds numbers involved 
is also quite broad. In this sense Zanoun et al.\ (2003) \nocite{zanounetal03} support $\kappa=0.37$ while 
Nagib et al. (2004) \nocite{nagibetal04} 
prefer 0.38, underlining  the influence of non-stationary conditions. Andreas et al.\ (2006) \nocite{andreasetal06} 
report a somewhat higher 0.387 $\pm\, 0.003$ based on experimental material
from the atmospheric boundary layer. However, in the Princeton superpipe a value of 0.421 has
been found for smooth flow at $Re \approx 3\cdot 10^7$ (Smits 2007)\nocite{smits07}, too.

Summarizing the hot debates of the past, a review 
by Jimenez and Moser (2007) \nocite{jimenezmoser07}  on wall turbulence states boldly: 
``The K{\'a}rm{\'a}n constant $\kappa \approx$ 0.4 is approximately universal.''
In practice this can hardly be distinguished from our primitive-theoretical value 
0.399 for $Re\rightarrow\infty$.

Nevertheless, doubts remain and the discussion by Talamelli \textit{et al.} (2009) is justified. 
They underline the limitations of the various existing superpipe 
facilities and propose a new one that will allow good spatial 
resolution even at very high Reynolds number. The new facility is 
the international effort CICLoPE and under construction in North Italy.

\subsection{Kinetic theory and spectra of turbulence\label{spectra}}

\noindent 
The above theory makes a precise prediction of von K{\'a}rm{\'a}n's constant $\kappa$.
Would it be possible to use the present kinetic picture of turbulence also to determine asymptotic values
for the Kolmogorov constants $C_K$ and $C_{\omega}$ in the spectral energy distributions over frequency and wave number?
With great probability the answer is no. Such an approach would demand that the lower bounds
$\lambda_u$ and $\omega_u$ in the energy integrals could be specified:
\begin{equation}
	K \;=\; C_K\,\varepsilon^{2/3} \int_{\lambda_u}^{\infty} k^{-5/3}\,dk\;=\; C_{\omega} \,\varepsilon \int_{\omega_u}^{\infty} \omega^{-2}\,d\omega\,. 
\end{equation}
These bounds are not simply connected with the ensemble-mean values of vorticity and length scale.

But it cannot be excluded that a dissipation theory of quasi-particles can be developed in analogy to a radioactive
decay chain, where large quasi-particles are broken by a collision into smaller ones etc. down to the
(infinitely small) Kolmogorov scale where they are annihilated. This could be imagined as a cup with a tiny
whole. It would keep all quasi-particles of size $>0$ in and let only those out which have zero size.
Possibly this approach could find some attention.

\subsection{Conclusions\label{conclusions}}

\noindent
The primitive picture of turbulence makes a minimum use of the Navier-Stokes equation.
Vortex properties are derived from the Euler equation and dissipaton is introduced through
an analogy to a birth and death process. Methodically our position is a neighbor of Maxwell's 
kinetic theory of real gases: 
\begin{itemize}

\item {\bf Real gas:} Free molecule flights and their collisions are strictly governed by inertia and short-range forces of a non-stationary
Coulomb field with moving point sources. Maxwell neglected the details of these interactions. He replaced them with 
geometrical and probability considerations and a statistical equilibrium assumption. He got the probability distribution 
of the molecular energies whereby advective, rotational and potential energy of the molecules 
could be treated as statistically independent. In an isolated real gas total energy 
and molecule number are conservation quantities. 

\item {\bf Turbulence:} Here neither energy nor particle number are conserved, but the flux of mechanical
energy into a volume element is in quasi-equilibrium with the heat flux across the boundaries. In vortex dipoles 
translation and rotation are coupled and statistically $not$ independent. Nevertheless one may apply 
Maxwell's central idea here as well. Although the movements of our quasi-particles 
are strictly governed by well-known rules (Navier and Stokes), we take them as unpredictable and replace them 
with geometry and probability arguments. Following Eward Lorenz (1960), 
such an approach stands today on much harder theoretical grounds than in Maxwell's times. \nocite{lorenz60}
\end{itemize}
We used only the following assumptions:
\begin{enumerate}
\item A turbulent fluid at $Re=\infty$ is a volume saturated with chaotically moving vortex-dipol tubes which are treated as quasi-particles.
\item A new-born quasi-particle is equipped with energy lost in from the mean flow
	formulated with the eddy-viscosity hypothesis, $-\langle u'\, w' \rangle = \nu\,S$. Its vorticity is governed by Tennekes' hypothesis. 
\item Advection of dipol tubes as quasi-particles follows simplest classical laws. 
\item If quasi-particles collide they are either scattered (50 \%) or annihilated/converted to heat (50 \%).
\item If scattered they perform a Brownian motion with Einstein's diffusivity in Prandtl-Kolmogorov formulation. 
\end{enumerate}

\noindent
{\bf Acknowledgments.} The author thanks Hartmut Peters in Seattle, Eckard Kleine in Hamburg and Michael Wilczek in M\"unster for their patience 
during turbulent discussions. He further thanks Rupert Klein in Berlin for still more patience, enduring interest, and substantial support.

\bibliography{kinetic_karman.bib}
\end{document}